\documentclass[twocolumn]{aastex631} % ,trackchanges
\usepackage{amsmath}
\usepackage{stackengine}
\usepackage{mathtools}
 \usepackage{gensymb}
 \usepackage{graphicx}
\usepackage{txfonts}
\usepackage[lf,enc=t1]{berenis}
\newcommand{\noop}[1]{}

\newcommand{\be}{\begin{eqnarray}}
\newcommand{\ee}{\end{eqnarray}}

\newcommand{\MSun} {\mbox{$M_{\odot}$}}
\newcommand{\Mstar}{\mbox{$M_{\star}$}}
\newcommand{\MEarth} {\mbox{$M_{\oplus}$}}

\shorttitle{Stellar mass-dependent disc lifetime distributions}
\begin{document}

\title{Disc lifetime distribution as a function of the mass of host star}

\author[0000-0002-5003-4714]{Susanne Pfalzner} 
\affiliation{J\"ulich Supercomputing Center, Forschungszentrum J\"ulich, J\"ulich, Germany}
%\affiliation{Max-Planck-Institut f\"ur Radioastronomie, Auf dem H\"ugel 69, 53121 Bonn, Germany}
\author[0009-0009-8603-9566]{Furkan Dincer} 
\affiliation{J\"ulich Supercomputing Center, Forschungszentrum J\"ulich, J\"ulich, Germany}
\affiliation{Department of Physics, University of Cologne, Cologne, Germany}
\author[0000-0003-2458-9756]{Nienke van der Marel} 
\affiliation{Leiden Observatory, Leiden University, Leiden, The Netherlands}
\author[0009-0001-5904-1742]{Frank W. Wagner} 
\affiliation{J\"ulich Supercomputing Center, Forschungszentrum J\"ulich, J\"ulich, Germany}

\correspondingauthor{Susanne Pfalzner}
\email{s.pfalzner@fz-juelich.de}

%\maketitle

\begin{abstract}
The lifetime of protoplanetary discs is a critical factor for planet formation. Although the mean disc lifetime provides an estimate of the typical period available for planet formation, it does not capture the substantial variability in individual disc lifetimes or their dependence on host star mass. This study addresses these limitations by deriving the disc lifetime distribution as a function of stellar mass. Our results reveal a pronounced mass-dependence. Performing a phenomenological fit using a Weibull distribution, we find the maxima of the distributions at $t_{max}^H =$~3.72~Myr for high-mass stars ($\approx$~1.00~--~3.00~\MSun) and $t_{max}^L =$~7.20~Myr for low-mass stars ($\approx$~0.01~--~0.20~\MSun) assuming an initial disc fraction of $f_{init} = 0.8$. All distributions are broad (typically \mbox{3.2~Myr~$< \sigma <$~4.7~Myr),} with the distribution for low-mass stars being somewhat broader. Our analysis indicates that not all stars are initially surrounded by a disc (60~\%~$< f_{init} <$~90~\% at cluster zero age), and that the initial disc fraction is even lower ($f_{init} \approx$~40~\%) for higher-mass stars. The potential mechanisms responsible for the observed spread and mass-dependence of disc lifetime distributions and initial disc fractions are discussed. Our primary aim is to demonstrate the methodology; more robust constraints will require improved data on mass-dependent disc fractions. Nevertheless, the derived mass-dependent disc lifetime distributions can already serve as a valuable input or a benchmark for planet-formation synthesis models.
\end{abstract}


\section{Introduction}
\label{sec:intro}

Discs around young stars contain gas and dust, the fundamental building blocks for forming planets. However, there is only a limited time to form planets from these materials. These discs disperse after a few million years through various processes \citep{Ribas:2014,Michel:2021,Pfalzner:2022,Brittain:2023}. After disc dispersal,  planet cores can no longer accrete gas, hindering existing cores from potentially developing into gas giants. Thus, a disc's lifetime may influence the types of planets that form and the overall architecture of the planetary system \citep{Ribas:2015,Pfalzner:2022}.

Given the importance of disc lifetimes, a lot of effort has been spent in establishing a typical disc lifetime \citep{Haisch:2001,Hernandez:2008,Fedele:2010,Ribas:2014,Pfalzner:2014,Richert:2018,Briceno:2019}. However, the idea of mean or median disc lifetimes has limitations. The lifespan of individual discs varies greatly. They can last less than 1~Myr or survive for more than 20~Myr \citep[e.g.,][]{Kennedy:2019,Kral:2020,Ronco:2021,Ballabio:2021}. In extreme cases, some last up to 40~Myr \citep{Silverberg:2016,Murphy:2018,Lee:2020}. Therefore, a disc lifetime distribution may better describe the situation \citep{Pfalzner:2024}.

In addition, the consistently lower disc fractions observed for high-mass stars indicate that these stars have significantly shorter typical disc lifetimes than discs around low-mass stars. Considering only the disc fraction in nearby associations ($<$~200~pc), \citet{Michel:2021} found disc lifetimes of 5~--~10~Myr instead of the often quoted 1~--~3~Myr. Generally, cluster selection seems to play a role when determining disc lifetimes. \citet{Pfalzner:2022} showed that young, distant clusters ($<$~5~Myr, $>$~200~pc) often dominated previous studies. Such clusters frequently suffer from limiting magnitudes, leading to an over-representation of higher-mass stars. As higher-mass stars disperse their discs earlier, the derived disc lifetimes are biased towards higher-mass stars. Based on the disc fractions observed by \citet{Luhman:2022}, \citet{Pfalzner:2022} estimated that the median disc lifetime for high-mass stars lies in the range of 2~--~5~Myr, compared to 5~--~10~Myr for low-mass stars.  Recently, the 5~--~10~Myr disc lifetime was confirmed by \citet{Polnitzky:2026} for discs in the Scorpius-Centaurus OB stellar groupings.
 The disc lifetime can be measured by different disc indicators, such as infrared excess, submillimeter observations, and H$\alpha$ emission. When referring to specific values, we consider the disc lifetimes derived from the disappearance of infrared excess.

As the median disc lifetime is a strong function of stellar mass, the disc lifetime distribution should also reflect this. Here, we extend the concept to disc lifetime distributions from the low-mass case to higher mass stars. We discuss the possible reasons for the stellar mass-dependence (Sect.~\ref{sec:mass_dependence}) and the large width of the disc lifetime distribution (Sect.~\ref{sec:width}).

\section{Method}
\label{sec:meth}

We build on the method suggested for low-mass stars in \citet{Pfalzner:2022b} and \citet{Pfalzner:2024}. We determine the disc lifetime distribution from the disc fractions in star clusters of different ages. The disc fraction $f_d(t)$ declines with time $t$ as the discs are dispersed as
\be
f_d(t) = f_d(0) - \int_0^t T_d dt \label{eq:fraction},
\ee
where $T_d$ denotes the individual disc lifetime and $f_d(0)$ the initial disc fraction. The disc lifetime distribution is then given by
\be
T_d(t) = - \frac{d}{dt} f_d(t) \label{eq:distribution}.
\ee
The principle is simple, but challenges arise due to uncertainties in cluster ages and disc fractions. These uncertainties must be appropriately accounted for when determining the disc lifetime distribution.

We test the quality of the fit $\Delta D$ by calculating the mean least-squares distance between the data points and the best-fitting curve. Thus,
\be
D = \sum_i  \frac{n_i}{\sum_i n_i} \sqrt{(\hat{t}_i - t_j)^2 + (\hat{f}_i - f_j)^2} \label{eq:fit_quality},
\ee
where $\hat{t}_i$ are the cluster ages, $\hat{f}_i$ the disc fractions; $t_j$ and $f_j$ represent the points on top of the distribution plotted from the initial parameters.

We pay particular attention to avoiding biases and selection effects, especially those sensitive to cluster selection \citep{Michel:2021,Pfalzner:2022}. The cluster sample used by \citet{Pfalzner:2024} included a considerable number of clusters with small membership sizes. Of the entire sample, only Upper Scorpius (Upp Sco) and Upper Centaurus Lupus/Lower-Centaurus-Crux (UCL/LCC) contain sufficient numbers of higher mass stars that separate disc fractions are meaningful. First, we use only these disc fractions to determine the mass-dependent disc lifetime distributions (see Sect.~\ref{sec:disc_lifetime}).

\section{Results}
\label{sec:res}

\begin{table*}
\caption{Disc fractions of the star‑forming regions Upp Sco and UCL/LCC for different spectral types and mass ranges.}
\centering\begin{tabular}{llcccccc}
\hline\hline
\multicolumn{1}{c}{Spectral type} &  Mass range, \MSun & \multicolumn{2}{c}{Disc fraction} & \multicolumn{2}{c}{No. stars}\\
\hline
& & $f^{UppSco}_{d}$ & $f^{UCL/LCC}_{d}$ & $N^{UppSco}_{e}$ & $N^{UCL/LCC}_{e}$\\
\hline
M9.75 -M3.75 & 0.01 -- 0.2  & 0.25$^{+0.05}_{-0.04}$          & 0.09$\pm$0.01             & 1301 & 2488\\

M3.5 -- K6 & 0.2 -- 1.00  & 0.18$^{+0.03}_{-0.02}$ & 0.03$\pm${0.01}           & 311 & 725\\

K5.5 -- B7 & 1.00 -- 3.0  & 0.05$^{+0.04}_{-0.03}$ & 0.007$^{+0.006}_{-0.004}$ & 76  & 452\\
\hline\hline
\bigskip
\end{tabular}
\label{tab:disc_fractions}
\end{table*}

\subsection{Disc lifetime distributions based on Upp Sco and UCL/LCC}
\label{sec:disc_lifetime}

First, we build on the work by \citet{Luhman:2022} and \citet{Luhman:2020}, where low-mass stars are defined as stars of spectral type M9.75~--~M3.75, intermediate-mass stars are of spectral type M3.5~--~K6, and high-mass stars are of spectral type K5.5~--~B7 (see Tab.~\ref{tab:disc_fractions}). These spectral types approximately correspond to the mass ranges 0.01~--~0.2~\MSun, 0.2~--~1.0~\MSun, and 1.0~--~3.0~\MSun, respectively. 
Table~\ref{tab:disc_fractions} also provides the disc fractions, $f_d$, for the star clusters Upper Scorpius (Upp Sco) and Upper Centaurus Lupus/Lower-Centaurus-Crux (UCL/LCC) given in \citet{Luhman:2020} and \citet{Luhman:2022}. In the next step, we add disc fractions from younger clusters to better restrain early disc dispersal. However, this comes at the cost of less consistency in the determination method.

The disc fraction is defined as the ratio of the number of stars showing infrared excess $N_e$ to the total number of stars $N$ observed in a given cluster. \citet{Luhman:2022} used photometry in three bands from WISE (W2, W3, W4) and three bands from Spitzer ([4.5], [8.0], [24]) to identify excess emission from discs among members of Upp Sco and UCL/LCC to classify the evolutionary stages of the detected discs. The evolutionary stage distinguishes between full discs, transitional discs, evolved transitional discs, and debris discs. They consider all of these classes except for the debris discs as primordial discs. Transitional discs have historically been marked as an evolved stage as the result of photoevaporative clearing \citep{Espaillat:2014}, but the last decade has demonstrated that many of the known transition discs with large cavities are, in fact, massive discs that have been cleared from warm dust due to other mechanisms, i.e., massive planets, and should therefore not be considered as an evolutionary stage that all discs go through \citep{vanderMarel:2023}, as the giant planet occurrence rate is typically $\lesssim$~10~--~20~\% \citep{Fernandes:2019}.

The number of stars included within each mass bin of a cluster affects the precision of the measured disc fraction and the statistical significance of the data point in the analysis. For this reason, we adjusted for statistical significance when analysing the data. The most statistically robust data points are those from Upp Sco and UCL/LCC because of their large numbers of stars. Nevertheless, all data points are subject to errors in disc fraction and cluster age. We address these uncertainties when deriving the disc lifetime distributions. The limited number of clusters older than 4~Myr reported in the literature -- including Upp Sco and UCL/LCC -- remains a notable issue, as it can introduce bias into conclusions. To ensure reliable data for high-mass stars, we limited the sample to clusters containing at least 200 stars summed over all mass bins.

Here, we used the bin classification by spectral class provided by \citet{Luhman:2020}. However, there exists a potential problem with using spectral types rather than stellar masses. Especially for young stars, the stellar mass depends (in combination with age) on the combination of luminosity and spectral type (temperature) following stellar isochrones \citep{Baraffe:2015}. However, in the absence of derived stellar masses for the majority of these samples, the spectral types are chosen as an approximation for the stellar mass bins.

\begin{figure}[ht]
  \centering
  \begin{minipage}[b]{0.45\textwidth}
    \centering
    \includegraphics[width=\textwidth]{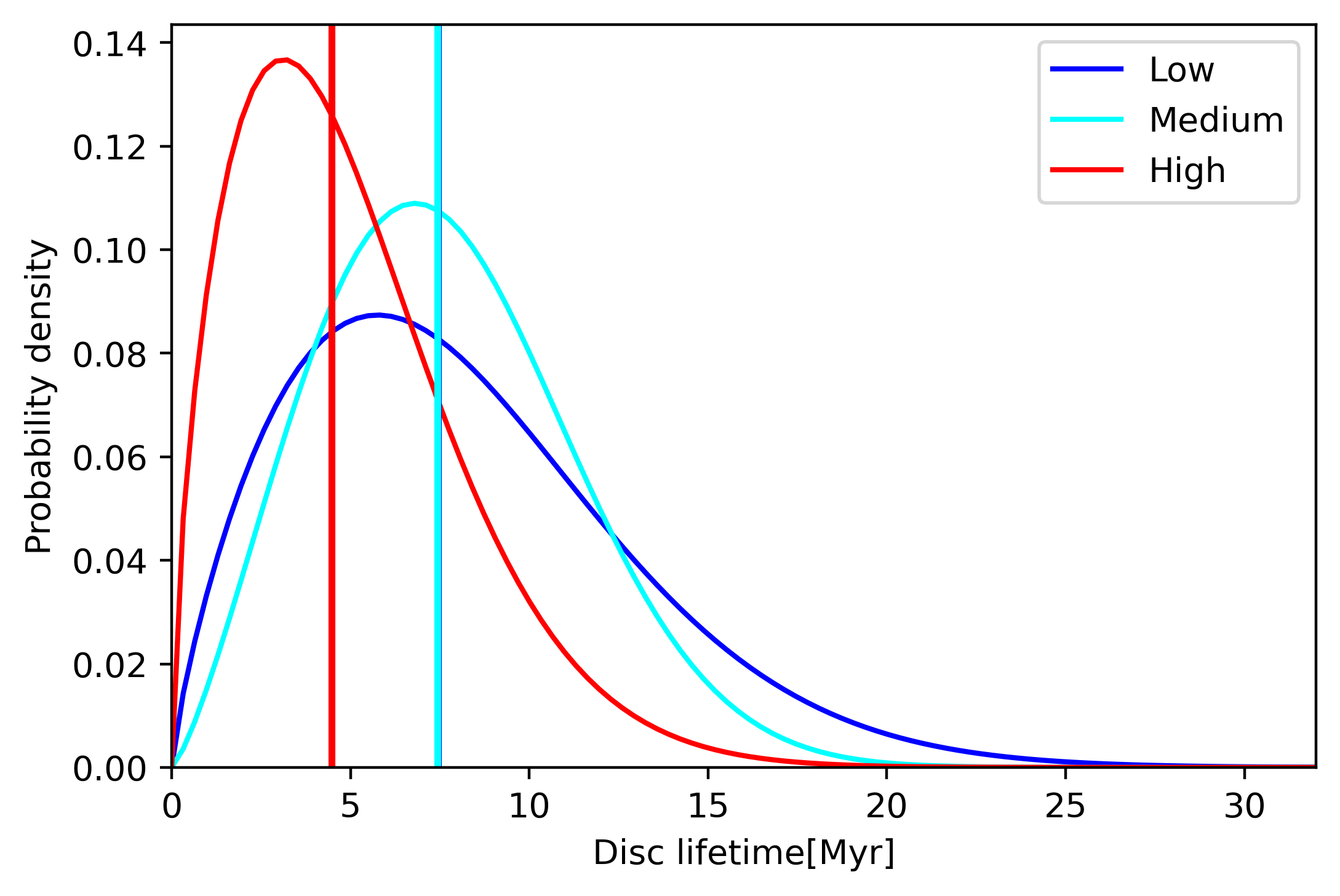}
    \textbf{(a)}
  \end{minipage}
  \begin{minipage}[b]{0.50\textwidth}
    \centering
    \includegraphics[width=\textwidth]{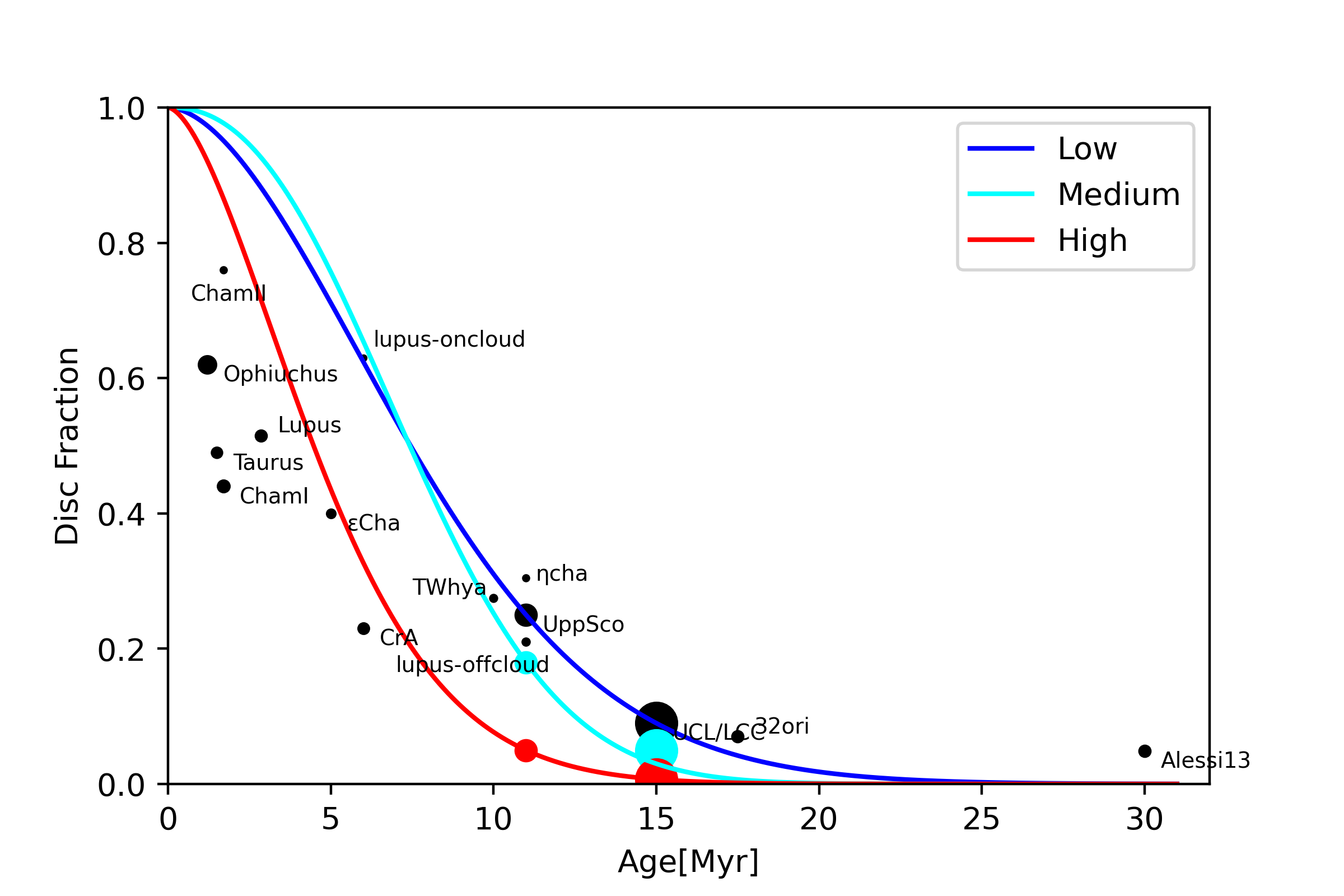}
    \textbf{(b)}
  \end{minipage}
 \caption{(a) Dependence of disc lifetime distribution on stellar mass. The best-fitting Weibull distribution is shown, assuming all stars start with a disc. We distinguish between low-, intermediate-, and high-mass stars as defined in the main text. Vertical lines indicate the median disc lifetime for each mass bin. (b) Corresponding survival disc fraction function. Observed disc fractions for star clusters within 200~pc are plotted for comparison (values and references in Tab.~\ref{tab:cluster_properties}). Data point sizes reflect sample sizes.}
\label{fig:distribution_mass_100} 
\end{figure}

\begin{figure}
 \centering
  \begin{minipage}[b]{0.46\textwidth}
    \centering
    \includegraphics[width=\textwidth]{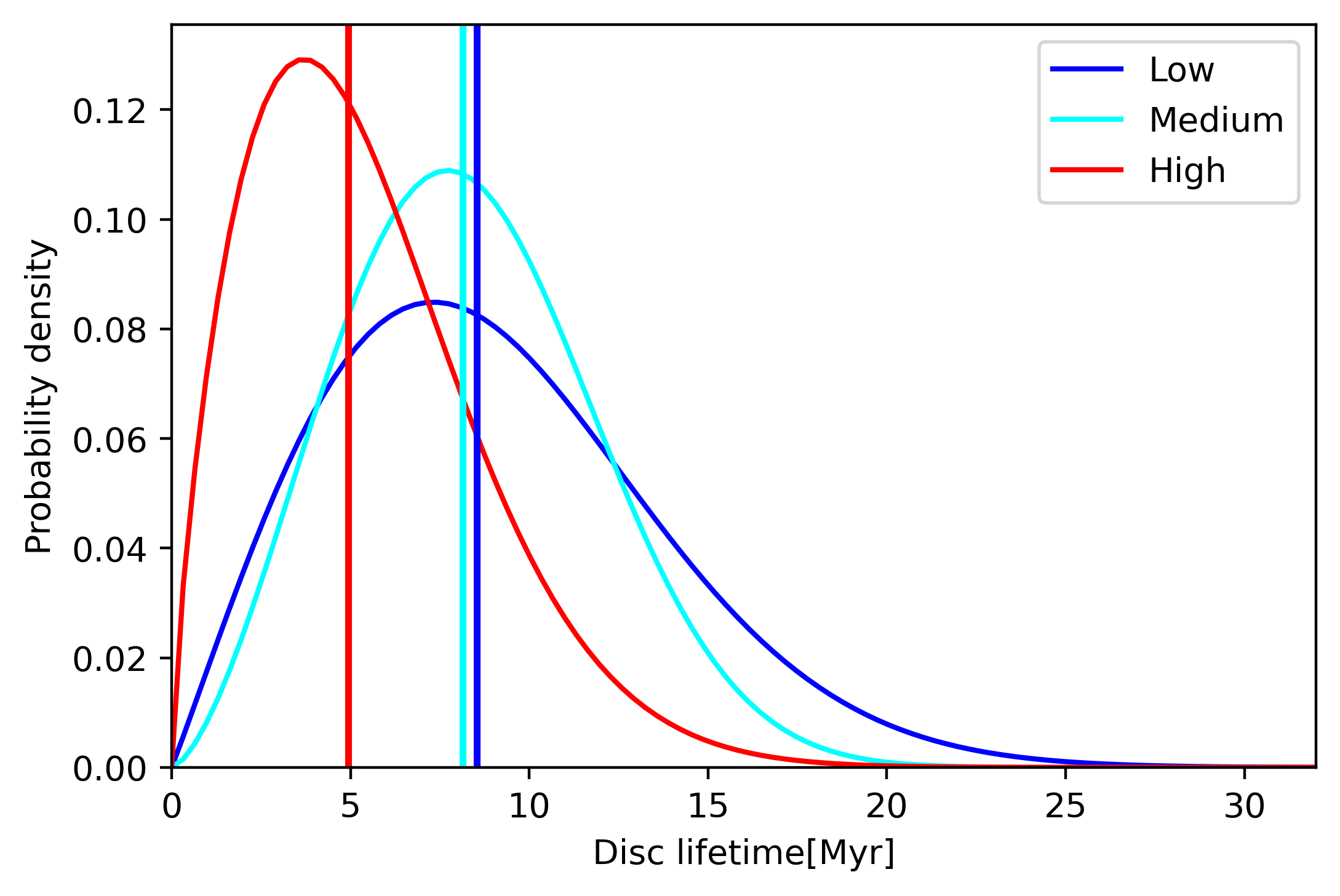}
    \textbf{(a)}
  \end{minipage}
  \begin{minipage}[b]{0.50\textwidth}
    \centering
    \includegraphics[width=\textwidth]{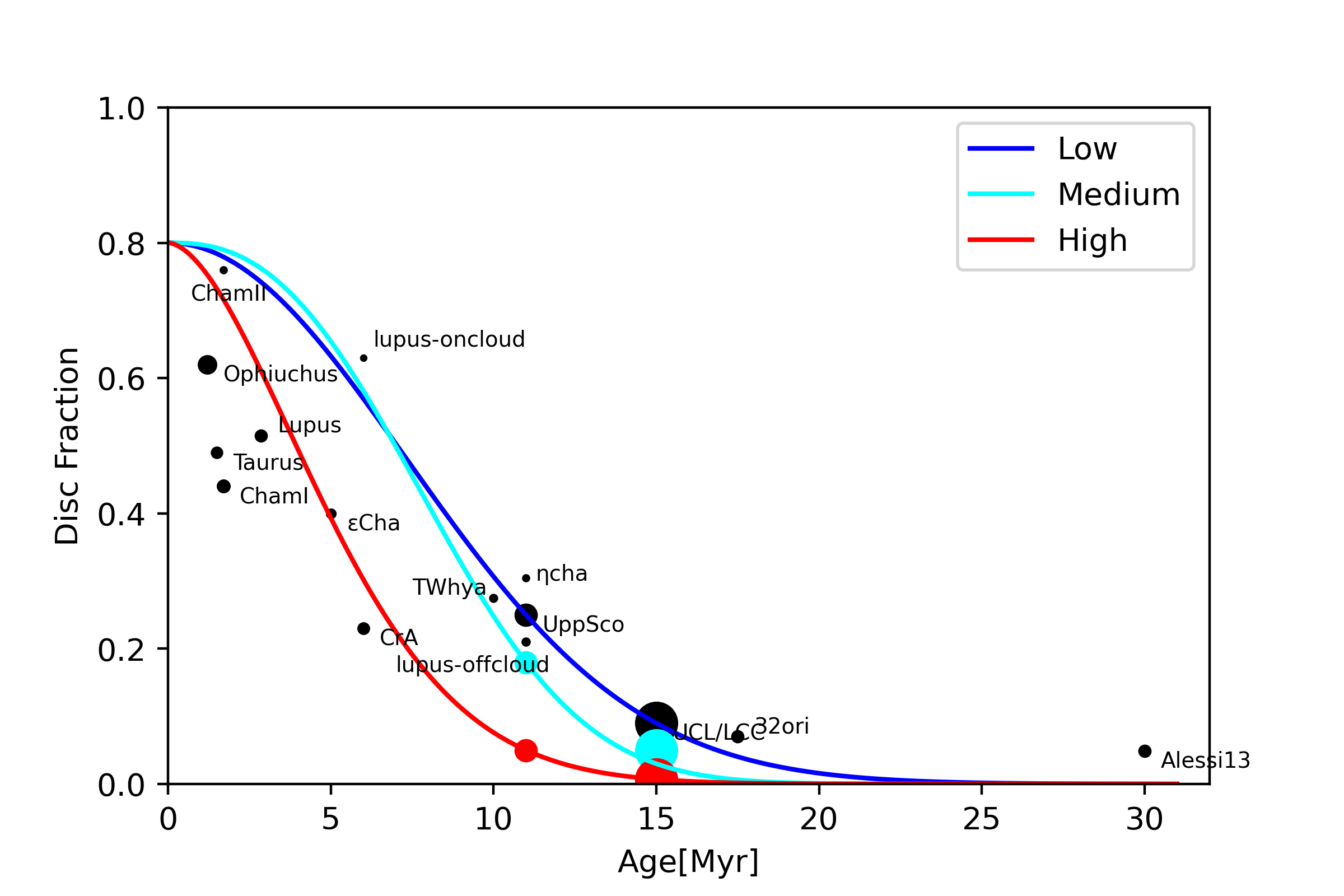}
    \textbf{(b)}
  \end{minipage}
\caption{Same as Fig.~\ref{fig:distribution_mass_100} but assuming only 80~\% of stars were initially surrounded by a disc.}
\label{fig:distribution_mass_80}
\end{figure}

\begin{table*}
\caption{Fit parameters of a Weibull-shaped distribution for three stellar spectral bins assuming two initial disc fractions. Column 1 indicates the spectral type, column 2 displays the mass bin, columns 3-7 show the case assuming an initial disc fraction of $f_{init} =$~100~\%, whereas columns 8-12 show the results for $f_{init} =$~80~\%. Given are the median disc lifetime $t_d$, the standard deviation of the distribution $\sigma (t_d)$, the quality of the fit $\Delta D $, and the maximum of the distribution $t_{max}$.}
\centering\begin{tabular}{llcccccccccc}
\hline\hline
& &\multicolumn{5}{c}{$f_{init}$ = 100~\%} &   \multicolumn{5}{c}{$f_{init}$ = 80~\%}\\
\hline
Spectral type & Mass bin, \MSun & Parameters & $t_d$ & $\sigma(t_d)$ & $\Delta D $ &  $t_{max}$ & Parameters & $t_d$ &  $\sigma(t_d)$  & $\Delta D $ &  $t_{max}$\\
\hline
M9.75 -- M3.75 & 0.01 - 0.2 &$k$ = 1.78, $\lambda$   = 9.15 &  7.45   &  4.73    &  0.0769  & 5.80  &$k$ = 2.03, $\lambda$  = 10.21   & 8.52    &  4.66  & 0.0523 & 7.20\\
M3.5 -- K6 & 0.2 - 1.0 &$k$ = 2.31 , $\lambda$   = 8.71  &  7.42  &   3.59  & 0.0836 & 6.87  &$k$ = 2.55 , $\lambda$  = 9.40   & 8.14  & 3.51 & 0.0579  &  7.66\\
K5.5 -- B7 & 1.0 - 3.0 &$k$ = 1.63 , $\lambda$   = 5.6  & 4.47   & 3.16    & 0.0672 & 3.22 &$k$ = 1.73 , $\lambda$  = 6.10   & 4.93   & 3.24 & 0.0478 & 3.72
\\
\hline\hline
\end{tabular}
\label{tab:fit_distributions_masses}
\end{table*}

It is not apparent what the shape of such a disc lifetime distribution should be. In \citet{Pfalzner:2024}, we systematically tested the ability of different distribution functions to represent a disc lifetime distribution of low-mass stars. We found that linear and exponential fits to the decline in disc fraction with cluster age lead to counterintuitive disc lifetime distributions. For example, exponential fits for the disc fraction result in exponential fits of the disc lifetime distribution, which implies that discs are most likely to be dispersed when they are very young ($<$~0.1~Myr) and there exists no maximum in the
distribution representative for a typical disc lifetime.  While a Gaussian provides a reasonable fit to the data, there exists a potential problem. It produces a cut-off of the distribution at $t =$~0~Myr, which can be interpreted as a considerable fraction of stars having negative disc lifetimes. Two conclusions exist: (i) the disc fraction distribution has a non-Gaussian shape, and (ii) the initial disc fraction is $\ll$~100~\%.

Generally, right-skewed distributions provide better fits to the data because they exhibit an initial steep decline and a long tail.  We fitted the observational data to different other types of distribution, namely, gamma, log-normal, Weibull, log-logistic, and beta, based on the disc fractions given in Tab.~\ref{tab:disc_fractions}. We also considered that these disc fractions have different statistical significances because the number of stars in the clusters differs considerably. It turned out that the Weibull distributions (Fig.~\ref{fig:distribution_mass_100}~a) provide good fits to the observational data. A Weibull has the following form:
\be
 T(t, k, \lambda) & = &  \frac{k}{\lambda} \left(\frac{t}{\lambda} \right)^{k-1} \exp \left(- \left(\frac{t}{\lambda} \right)^k \right),
\ee
where $k$ and $\lambda$ are the fit parameters. We determine the parameters $k$ and $\lambda$ that provide the best fit to the observational data (Fig.~\ref{fig:distribution_mass_100}~b). The values for the different mass bins can be found in Tab.~\ref{tab:fit_distributions_masses} together with the corresponding median disc lifetime $t_d$, the standard deviation $\sigma$, the fit quality $\Delta D$, and the maximum of the distribution $t_{max}$. A key limitation lies in the large error bars associated with disc fractions and cluster ages (see Table \ref{tab:cluster_properties} and Fig. \ref{fig:distribution_mass_error} in the Appendix) — uncertainties that significantly hamper the fitting procedure. Tighter observational constraints on these quantities are critically needed. Nevertheless, this study adopts the available values as given, prioritizing the demonstration of the methodology over precise parameter fitting.

In the following, we use the Weibull distribution. One reason is that this is consistent with the distribution we use in \citet{Pfalzner:2024}. The fit is purely phenomenological. For any right-skewed distribution, there exists a peak and a median value. The peak values correspond to the time when the largest number of stars lose their discs, whereas the median disc lifetime refers to the typical disc lifetime. To illustrate the difference, for an exponential function, the peak would be at $t =$~0~Myr.

We find that the distribution of intermediate-mass stars is only moderately different from the distribution we identified for low-mass stars in \citet{Pfalzner:2024}. The median disc lifetime for intermediate-mass stars is 7.42~Myr, compared to 7.45~Myr for low-mass stars. Due to the high uncertainty in disc fractions and cluster ages, it is possible to consider both low-mass and intermediate-mass stars to have a common distribution.

By contrast, there is a significant shift towards shorter disc lifetimes in the distribution of high-mass stars. For high-mass stars, the maximum in the distribution is at $t^H_{max} =$~3.22~Myr, whereas the corresponding value for low-mass stars is $t^L_{max} =$~5.80~Myr. 
Due to the distributions being right-skewed, the mean disc dispersal times are higher -- for high-mass stars, the median lifetime is $t_d^H =$~4.47~Myr compared to $t_d^L =$~7.45~Myr. These values are indicated as vertical lines in Fig.~\ref{fig:distribution_mass_100}~a.

In rare cases, protoplanetary discs have been found surrounding stars that are 20~Myr old and older. However, the host stars are more or less exclusively low-mass \mbox{$M$-type} stars. This situation is also reflected in our distributions for low-mass and high-mass stars. For low-mass stars, 95~\% of stars have lost their discs by an age of 17.5~Myr. For high-mass stars, this value drops to 11.5~Myr. As such, there would be only a 0.04~\% chance of finding a 20~Myr old high-mass star still being surrounded by a protoplanetary disc.

Figure~\ref{fig:distribution_mass_100}~b illustrates the corresponding functions of the disc fraction versus cluster age. In addition to the Upp Sco and UCL/LCC data points, the figure includes observed disc fractions for clusters located within 200~pc (see Tab.~\ref{tab:cluster_properties} for values and references). For these clusters, disc fractions are not separated by stellar mass due to the limited number of high-mass stars in the small samples. Consequently, most cluster members are of relatively low mass. The size of each data point reflects the number of stars in the corresponding sample.

We had already noted in \citet{Pfalzner:2024} that disc fractions at young ages ($<$~2~Myr) are not as high as one would expect if discs initially surrounded all stars. \citet{Michel:2021} also concluded that the initial disc fractions in the young regions should be considerably below 100~\%. They obtained much better fits to their data assuming initial disc fractions between 65~\% and 85~\%. They suggested that disc destruction in close binaries ($<$~40~au) could be the reason. In Taurus, at least 15~\% to 20~\% of the discs could have been destroyed or prevented from forming due to such close binaries \citep{Kraus:2012}. Close binaries have been observed to lead to less and smaller mm-dust emissions \citep{Harris:2012,Aleson:2019}. 
Other reasons may be that some stars are born without a disc, have lost their disc rapidly \citep{Richert:2018}, or age spreads in clusters \citep{Kenyon:1995,Terwisga:2022}. Even if all stars were initially surrounded by a disc, at cluster age $t =$~0~Myr, the oldest stars in the cluster may already be several Myr old and have lost their discs, leading to $f_d(0) <$~100~\%.

The advantage of using data points from \citet{Luhman:2020} and \citet{Luhman:2022} is that the disc fractions of low‑mass and high‑mass stars were derived self-consistently. However, deriving distributions from just two data points per mass bin is non-ideal. Next, we derive disc lifetimes by adding additional data points.

\begin{figure}[ht]
  \begin{minipage}[b]{0.44\textwidth}
    \centering
    \includegraphics[width=\textwidth]{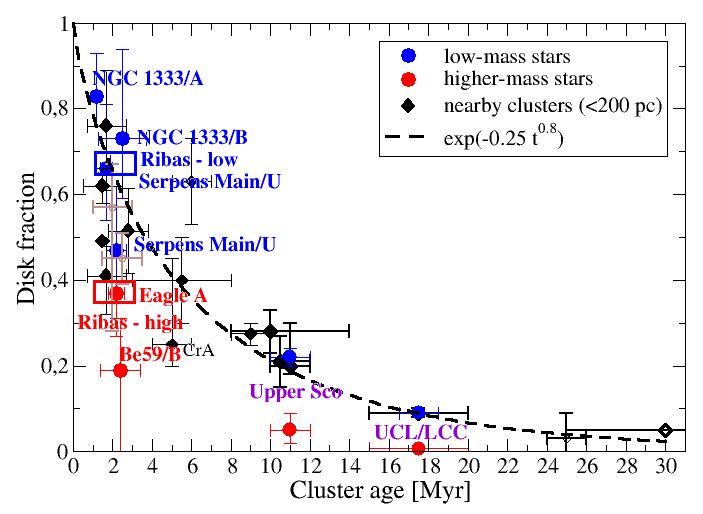}
    \textbf{(a)}
  \end{minipage}
  \begin{minipage}[b]{0.44\textwidth}
    \centering
    \includegraphics[width=\textwidth]{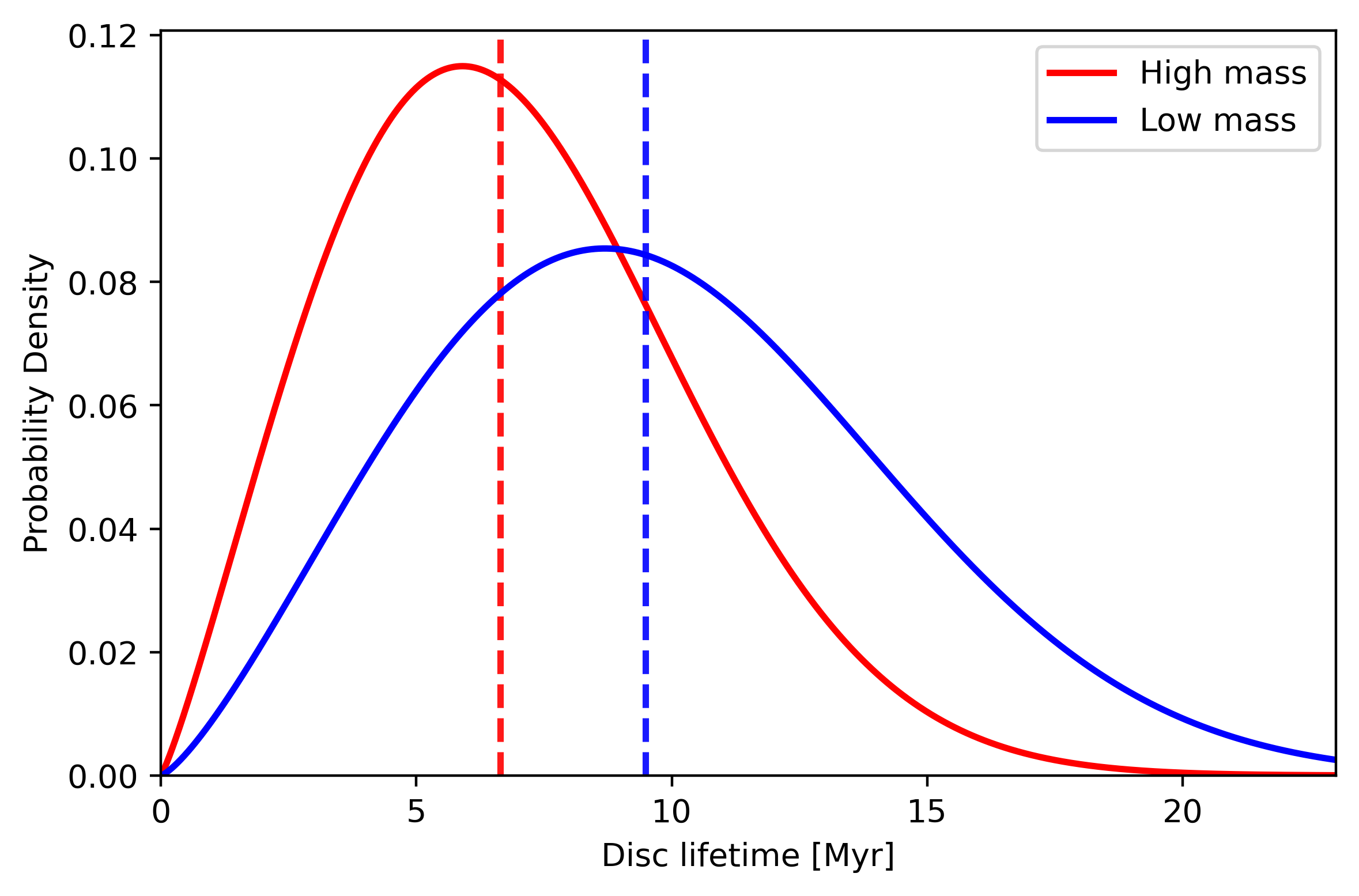}
    \textbf{(b)}
  \end{minipage}
  \begin{minipage}[b]{0.44\textwidth}
    \centering
    \includegraphics[width=\textwidth]{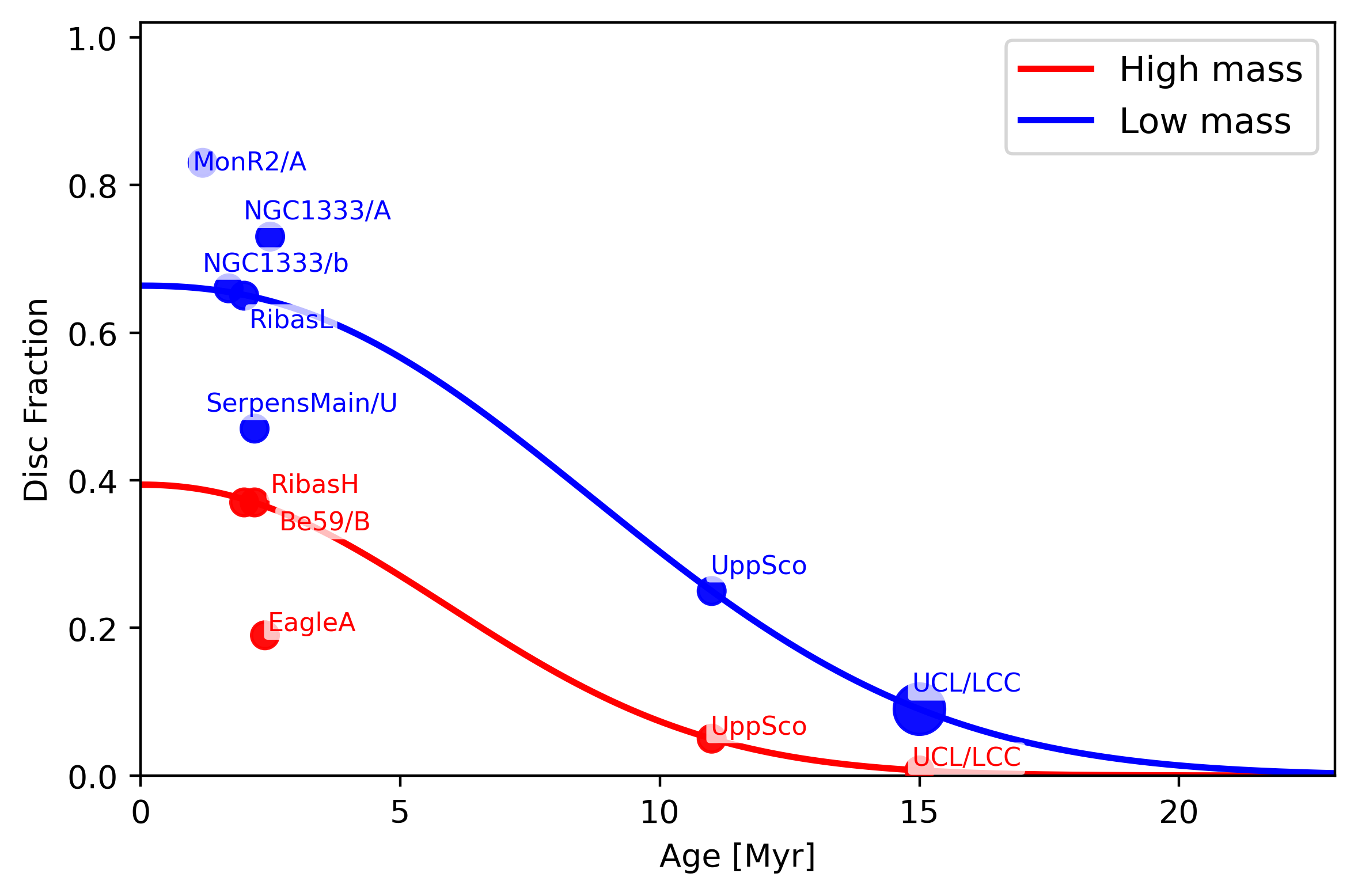}
    \textbf{(c)}
  \end{minipage}
\caption{(a) Observed disc fractions for low-mass (blue) and high-mass (red) stars used for deriving the disc lifetime distribution. (b) and (c): Same as Fig.~\ref{fig:distribution_mass_100} but taking the disc fraction of young high-mass stars from \citet{Ribas:2015} and \citet{Richert:2018}
into account. Here, we left the initial disc fraction as a free parameter.}
\label{fig:distribution_mass_free}
\end{figure}

\subsection{Disc lifetime distributions adding younger clusters}
\label{sec:disc_lifetime_young}

The challenge is that there are very few data available on disc fractions for high-mass stars. Next, we include three data points extracted from \citet{Ribas:2015} and \citet{Richert:2018}. We did not include them in the previous plots because they span somewhat different mass bins and the disc fraction methodology differs. However, they clearly support the trend of a lower disc fraction for higher-mass stars at cluster ages $<$~3~Myr. \citet{Ribas:2014} obtained a disc fraction of 35 to 40~\% for host stars with $\Mstar >$~2~\MSun, compared to 60 to 70~\% for host stars with $\Mstar <$~2~\MSun. \citet{Richert:2018} determined the disc fractions of 69 young clusters ($<$~5~Myr) located at distances of 0.2~--~3.6~pc. They provide information on the mass cut-off and the mean stellar mass. Here, we regard clusters that have a mass cut-off $>$~0.75~\MSun\ and a mean mass of host stars $>$~1.0~\MSun\ to provide disc fractions that are representative of higher-mass stars. Clusters with a mass cut-off $<$~0.1~\MSun\ and a mean mass of host stars $<$~0.3~\MSun\ are used to represent low-mass stars (see Tab.~\ref{tab:disc_fractions_other}).

Although these distributions benefit from a greater number of data points, the derivation of disc fractions lacks methodological consistency. In both instances, the quantity and quality of the data remain limited. The primary objective of this study is to illustrate the concept and significance of mass-dependent disc lifetime distributions. With the acquisition of additional data, it will be possible to constrain these distributions more effectively.

\begin{table*}
\caption{Disc fractions for young star clusters.}
\centering\begin{tabular}{lllcccccc}
\hline\hline
\multicolumn{1}{c} {Bin} &{Source} & Cluster  & $N_{e}$ &  Age, Myr  & \multicolumn{3}{c}{Stellar mass, \MSun} & \multicolumn{1}{c}{Disc fraction}\\
\hline
         &     &     &       &  & range & cut & mean \\
\hline
High-mass stars &
\citet{Ribas:2015}   & several & 36 & $2^{+1}_{-1}$ & 1 -- 3 &  > 2  &   -   &    35 -- 40         \\
&\citet{Richert:2018} & Eagle A & 16 & $2.4^{+1}_{-1}$ & - &     0.95    & 1.24 & 0.19$^{+0.12}_{-0.19}$  \\
&\citet{Richert:2018} & Be59/B  & 95 & $2.2^{+0.4}_{-0.4}$ &  -  &  0.75   &    1.59        &  0.37$^{+0.09}_{-0.10}$   \\
\hline
Low-mass stars &
\citet{Ribas:2015}   & several & 176 & $2^{+1}_{-1}$ & 1 -- 3 &  < 2 &   -   &    60 -- 70               \\
&\citet{Richert:2018} & NGC 1333/A & 22 & $2.5^{+1.2}_{-1.2}$ &  - &  0.09         & 0.17  & 0.$73^{+0.21}_{-0.14}$  \\
&\citet{Richert:2018} & NGC 1333/b & 44 & $1.7^{+0.3}_{-0.3}$ & - &    0.09          & 0.18 & 0.66$^{+0.15}_{-0.12}$   \\
&\citet{Richert:2018} & Serpens Main/U & 36 & $2.2^{+0.5}_{-0.5}$ & - &    0.09          & 0.19 & 0.47$^{+0.15}_{-0.16}$  \\
&\citet{Richert:2018} & Mon R2/A & 78 & $1.2^{+0.1}_{-0.1}$ & - & 0.09 & 0.28 & 0.83$^{+0.10}_{-0.07}$   \\
\hline\hline
\end{tabular}
\label{tab:disc_fractions_other}
\end{table*}

\begin{table}
\caption{Fit parameters of a Weibull-shaped distribution for low-mass and high-mass stars using the extended disc fraction set. For the definitions of low- and high-mass stars, see the main text. Column 1 indicates the stellar mass type, column 2 the fit parameters, column 3 the median disc lifetime $t_d$, column 4 the standard deviation of the distribution $\sigma (t_d)$, column 5 the quality of the fit $\Delta D$, and column 6 the maximum of the distribution $t_{max}$.}
\centering\begin{tabular}{lcccccccccc}
\hline\hline
% &\multicolumn{5}{c}{low-mass stars} &   \multicolumn{5}{c}{high-mass stars}\\
%\hline
Mass & \multicolumn{1}{l}{Parameters} & $t_d$ & $\sigma(t_d)$ & $\Delta D $   &$t_{max}$ \\
\hline
low &$k$ = 2.31 , $\lambda$   = 11.12   &   9.48  &  4.53  &  0.0636  &  8.69 \\
high
 &$k$ = 2.16, $\lambda$   = 7.87    &   6.64  &  3.4  &  0.0357  &  5.91 \\
%\\
\hline\hline
\end{tabular}
\label{tab:fit_distributions_young}
\end{table}

\subsection{Initial disc fraction}
\label{sec:init_fraction}

The distributions themselves, as well as the median and maximum values, are influenced by the assumed initial disc fraction. Figure~\ref{fig:distribution_mass_80} demonstrates this sensitivity. For an assumed initial disc fraction $f_d(0) =$~80~\%, the distribution extends to slightly longer disc lifetimes with the median lifetime $t_d =$~4.93~Myr and the maximum lifetime $t_{max} =$~3.72~Myr for high-mass stars. If we decrease the initial disc fraction to $f_d(0) =$~65~\%, the shift would be even greater with $t_d =$~5.39~Myr and $t_{max} =$~4.39~Myr.

Visual inspection hints at the $f_d(0) =$~80~\% case, providing a better fit to the data than the $f_d(0) =$~100~\% case. The improvement in the fit is quantified in the quality parameter. All $\Delta D$ values are lower for the $f_d(0) =$~80~\% case, with the fit to the high-mass stars having the smallest error. This result agrees with the results of \citet{Michel:2021}, which find that their fits of exponential functions improve if they assume $f_d(0) \ll$~100~\%.

Next, we investigate the best fit when leaving the initial disc fraction as a free parameter. In addition to Upp Sco and UCL/LCC, the disc fractions of the clusters given in Tab.~\ref{tab:disc_fractions_other} are used to find the best fit. Figure~\ref{fig:distribution_mass_free} shows the resulting distributions. For the corresponding parameters of the distributions, see Tab.~\ref{tab:fit_distributions_young}. In both cases, the initial disc fraction is $\ll$~100~\% -- for low-mass stars it is $f^L_{init} =$~66.4~\% and for high-mass stars it is even lower, namely, $f^H_{init} =$~39.4~\%. If the initial disc fractions are really that low, this would also affect the distributions, shifting the median disc lifetime to larger values. The lifetime distribution for low-mass stars would be $t^L_d =$~9.48~Myr with $\sigma(t^L_d) =$~4.53~Myr and for high-mass stars $t^H_d =$~6.64~Myr with $\sigma(t^H_d) =$~3.4~Myr (Fig.~\ref{fig:distribution_mass_free}~b and c).

A lower initial disc fraction means a shift of the reference frame for all disc fractions at later times. For example, a 25~\% disc fraction at 5~Myr, for an initial disc fraction that means 75~\% of discs are removed, whereas for a 50~\% disc fraction, only 50~\% of discs are removed at 5~Myr. Naturally, the disc lifetime of the 100~\% initial disc fraction is shorter than for the 50~\% disc fraction. In terms of processes, it would mean: If the real initial disc fraction is significantly less than 100~\%, the viscous evolution must be slower, or the photo-evaporation and winds weaker than would be assumed for $f_{init}=$~100~\%.

We find strong indications that the initial disc fraction may depend on stellar mass. This situation may also be rooted in stellar binarity. Observations show that the binarity/multiplicity of stars is a strong function of stellar mass \citep{Duchene:2013,Sana:2014}. \mbox{$M$-type} stars (0.1~\MSun\ $< \Mstar <$~0.5~\MSun) have considerably lower multiplicity fraction ($f_m =$~0.26) than \mbox{$O$-type} stars ($f_m =$~0.8) \citep{Duchene:2013,Offner:2023}. A possible consequence would be that the larger binary fraction among high-mass stars may lead to a lower initial disc fraction, as fewer discs are formed or more discs are destroyed on very short time scales ($\ll$~1~Myr) due to the binary companion.

Cluster ages carry significant uncertainties depending on the adopted stellar evolutionary models and age-dating techniques, so we account for errors in age and disc fraction during fitting. According to \citet{Pecaut:2016}, table~12, the intrinsic age spread adopted for UCL/LCC is 15~$\pm$~5~Myr, with a median age of 16~$\pm$~2~Myr, widely used as a benchmark. We run the fit with both values using our bounded LS fitting algorithm. Given the large uncertainty in the 15~Myr age ($>$~30~\%), a full Bayesian analysis would be warranted, but it is beyond the scope of this paper; therefore, these results serve as a first assessment of robustness with respect to age uncertainties. For 16~Myr, we obtain ($t_d =$~9.01~Myr, $f_{init} =$~0.6840~$\pm$~0.0194) for the low-mass case and ($t_d =$~5.80~Myr, $f_{init} =$~0.4177~$\pm$~0.0091) for the high-mass case. For 15~Myr, we obtain ($t_d =$~9.43~Myr, $f_{init} =$~0.6704~$\pm$~0.0199) and ($t_d =$~6.62~Myr, $f_{init} =$~0.3948~$\pm$~0.0081), respectively. The trend is robust: low-mass stars have longer median disc lifetimes ($\sim$~68~\% initial disc fraction) and high-mass stars shorter ones ($\sim$~41~\%).

Planet statistics support such a scenario: In single-star systems, planets occur 4.5$\pm${3.2}, 2.6$\pm${1.0}, and 1.7$\pm${0.5} times more frequently than in binary systems when a stellar companion is present
at a distance of 10, 100, and 1000~au, respectively \citep{Wang:2014}. Particularly in very close binary systems ($<$~1~au), planets around only one of the two stars are fully suppressed \citep{Moe:2021}. About 40~\% of solar-type primaries do not host close planets due to suppression by close stellar companions \citep{Kraus:2016}, likely because stellar companions disrupt or suppress the formation and/or survival of protoplanetary discs \citep{Kraus:2012,Barenfeld:2019}.

\subsection{Interpretation of distributions}
\label{sec:dist}

\begin{figure}[ht]
\includegraphics[width=0.47\textwidth]{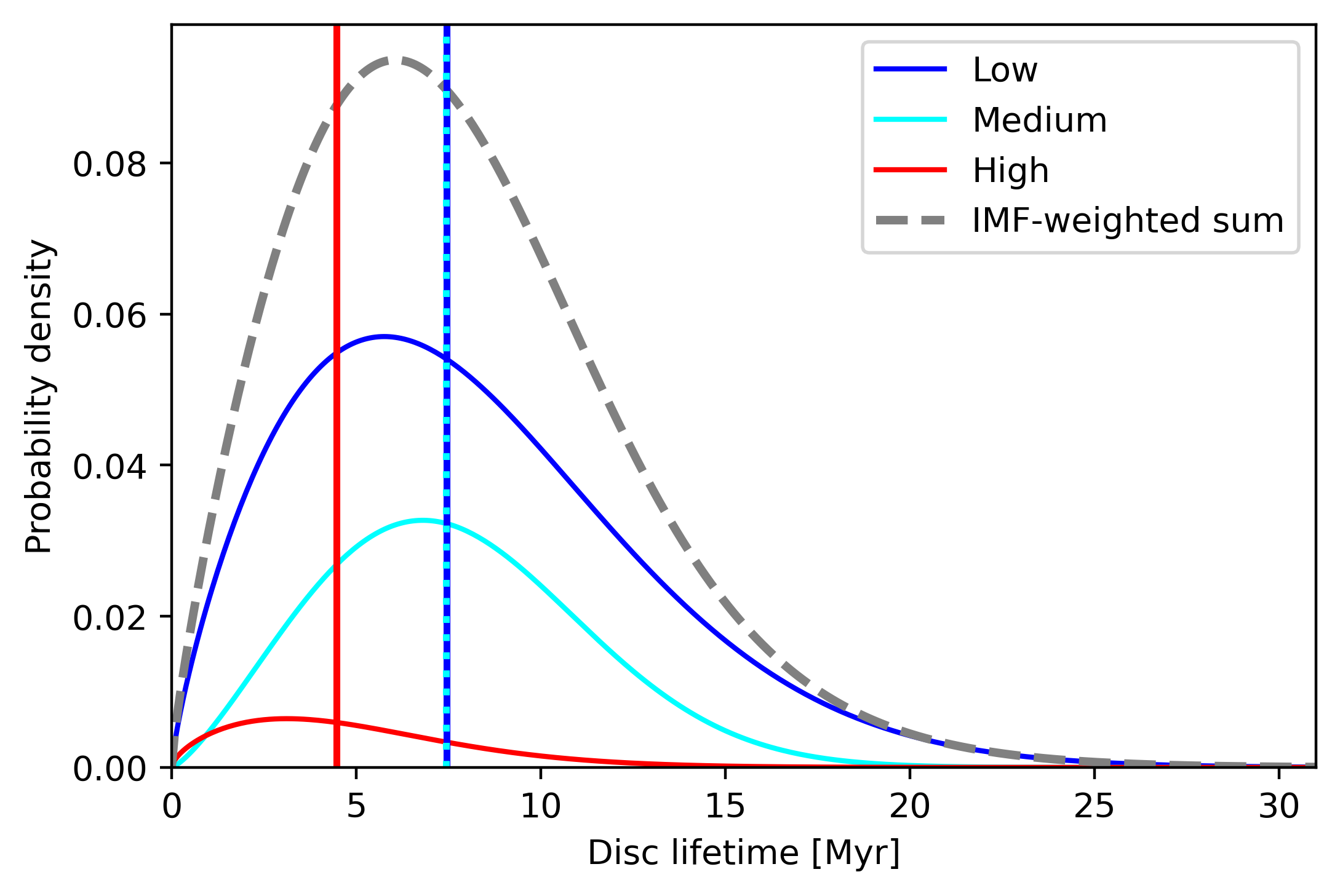}
\includegraphics[width=0.48\textwidth]{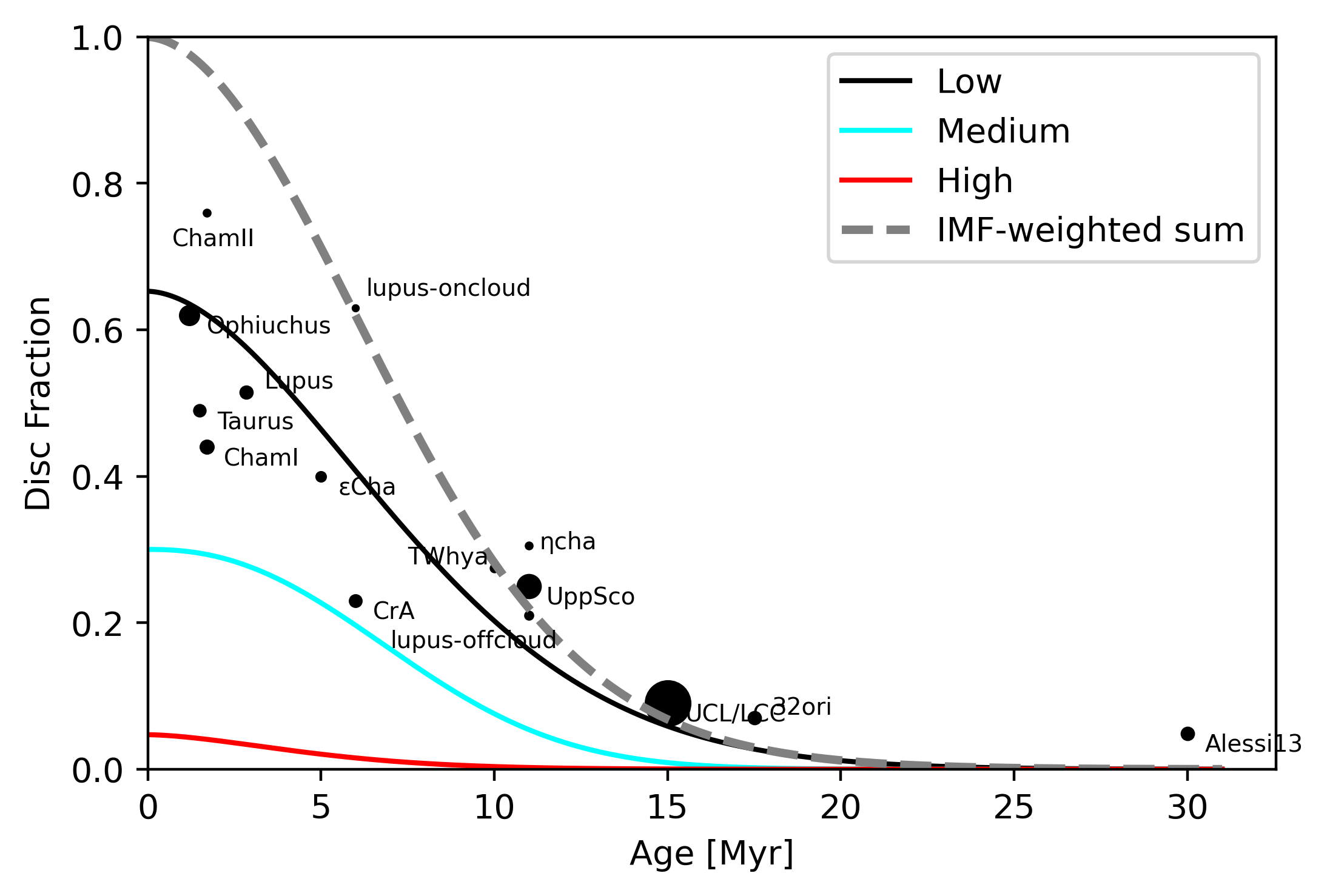}
\caption{Relative contributions to the total disc lifetime distribution from the low-, intermediate-, and high-mass stars. Contributions are calculated according to the relative fractions of the different mass bins assuming a Kroupa IMF and normalized over three mass bins.}
\label{fig:distribution_schematic}
\end{figure}

So far, we have discussed the disc lifetime distributions for the different mass bins separately.
The next step is to construct the overall disc lifetime distribution, taking into account the initial mass function (IMF). Here, we use the IMF, suggested by \citet{Kroupa:2001}, which is given by
\begin{equation}
    \label{eq:imf_kroupa}
    \xi (m) \propto \begin{cases} 
      m^{-0.3} & 0.01\leq m/M_{\odot} < 0.08, \\
      m^{-1.3} & 0.08\leq m/M_{\odot} < 0.5, \\
      m^{-2.3} & 0.5\leq m/M_{\odot},  
   \end{cases}
\end{equation}
with the number of stars $dN$ and with masses $m$ in the range to $m + dm$, $dN= \xi(m) dm$.

The overall disc lifetime distribution can be constructed from the individual mass bins by determining the relative portions of stars in the three mass bins. Specifically, 65.3~\% of the stars belong to the low-mass bin, 30.0~\% to the intermediate-mass bin, and only 4.7~\% to the high-mass bin. The total disc lifetime distribution results from summing the contributions from each bin, weighted by their relative weights. Figure~\ref{fig:distribution_schematic} illustrates this distribution, showing that the contributions from low-mass and intermediate-mass stars dominate the total disc lifetime distribution. We used the combined disc lifetime distribution to determine the fraction of stars with discs that persist for more than 5~Myr which is $70.74 \ \%$. 
The total IMF-weighted by the mass bin's contributions is described by the following parameters: $k$ = 1.71 and $\lambda$ = 8.19. The median disc lifetime is $t^{ws}_d$ = 7.3~Myr; it reflects the dominance of low- and medium-mass stars and is nearly identical to the medium disc lifetime of these stars.

The results presented above prompt two immediate questions:
\begin{itemize}
\item \emph{Why do discs tend to have shorter lifetimes around more massive stars?} 
\item \emph{Why are all disc lifetime distributions so wide?} 
\end{itemize}

In the following, we discuss various models and compare their findings with our results on the distribution of disc lifetime. Although in reality both questions are likely connected, we discuss them separately for the sake of simplicity of the argument.

\section{Possible Reasons for mass-dependence}
\label{sec:mass_dependence}

The disc masses of high-mass stars are, on average, higher than those of low-mass stars \citep[e.g.,][]{Pascucci:2016}. If the accretion rates were independent of stellar mass, higher-mass stars would live longer. However, observations show the opposite: despite higher disc masses, high-mass stars disperse their discs 2 to 4 times faster. Although several survey studies found elevated accretion rates around higher mass stars (e.g., figure~4 in \citet{Manara:2023}), the increased accretion rate is on its own insufficient to account for the large difference in disc lifetimes.
There could be several reasons for the short disc lifetimes of higher mass stars. The disc dispersal mechanisms may be much more efficient around high-mass stars, the disc structure may play a role, or differences in disc composition could influence the dispersal process. Alternatively, differences in the planet-formation process itself may lead to premature disc dispersal, mostly around high-mass stars. Naturally, a combination of these effects could also be at work. In the following, we discuss various models and compare their findings with our results on the distribution of disc lifetimes.

An important constraint from observations is that the turnover point from short-lived to long-lived discs appears somewhere at a host star mass of 0.5~--~2~\MSun\ \citep{Ribas:2015, Pfalzner:2022}. Unfortunately, this transition point is not yet better constrained.

\subsection{Disc dispersal mechanisms}
\label{sec:dispersal_mechanisms}

Several numerical models have been developed to reproduce the observed properties of protoplanetary disc populations using simplified physical prescriptions. The evolution of protoplanetary discs is studied primarily using two theoretical frameworks: the viscous accretion disc model introduced by \citet{Lynden:1974} and the magnetically driven disc wind model proposed by \citet{Blandford:1982}. Accretion onto the star occurs because of the transfer of angular momentum within the disc, facilitated by turbulent stresses. It also occurs when magnetohydrodynamic (MHD) disc winds pull gas out of the disc. MHD winds, along with those generated by the photoevaporation of the disc's surface layers, can transport gas into interstellar space. Despite extensive research, the relationship between the mechanisms of disc dispersal and stellar mass is still not fully understood.

Especially, when including stellar evolution, such studies usually find a dependence on the host star mass \citep{Kunitomo:2021, Ronco:2024}. FUV radiation can play a major role in driving internal photoevaporation, which can strongly influence disc dispersal. However, they recover the mass trend only partly. Especially, finding the observed, relatively abrupt, shortening of disc lifetime somewhere in the interval 0.5--2.0~\MSun\ \citep{Pfalzner:2022} seems a particular challenge.
Many studies concentrate on stars with $>$~1~\MSun, excluding stars of the lowest mass ($<$~0.5~\MSun) that dominate the IMF. These studies often find a decrease in disc lifetime in the mass range 2~\MSun\ $< \Mstar <$~3~\MSun\ \citep{Kunitomo:2021,Komaki:2023}. However, contrary to observations, simulations indicate that lower-mass stars ($<$~1~\MSun) tend to disperse their discs more quickly \citep{Komaki:2023}.

Simulations also yield varying conclusions about how external photoevaporation affects disc dispersal. Although \citet{Coleman:2022} and \citet{Gomez:2025} find it essential to include external photoevaporation, \citet{Emsenhuber:2023} believe that it is not required. They find that their models better match the properties of discs in low-mass star-forming regions when they assume more massive, compact discs without external photoevaporation. \citet{Coleman:2022} and \citet{Gomez:2025} argue that ongoing star formation is needed to explain the observed disc fractions related to the age of clusters.

\citet{Wilhelm:2022} concentrated on discs with ages that far exceed the typical protoplanetary disc lifetime. They found shorter \emph{maximum} lifetimes for stars with masses $>$~0.8~\MSun. They argue that the disc lifetimes should be inversely proportional to the host star's mass -- a prediction that observations could test if the dependence of disc lifetime on stellar mass were known in greater detail. So far, observations seem to point more to a turnover point than to a steady decline \citep[see figure~3 in][]{Pfalzner:2022}. Unfortunately, the location could so far only be vaguely defined in the range 0.8~--~1.8~\MSun.

In summary, there is no definitive answer yet, which process describes the disc dispersal better. 
Although these numerical simulations provide valuable insight, the results are susceptible to the choice of initial conditions, numerical methods, and model parameters, such as the magnetic field strength \citep{Coleman:2022}. Therefore, it is challenging to make stringent predictions that could rule out models.

\subsection{Disc structure}
\label{sec:structure}

Alternatively, the disc's structure may influence its lifetime. 
There seems to be a relation between the presence of large scale dust substructures and the dust mass evolution in discs \citep{Marel:2021,Savvidou:2025} it was found that at all ages ($\leq$~10~Myr), structured dust discs (interpreted as discs with pressure bumps) retained high dust masses, while the dust mass of non-structured discs (compact discs) decreased over time, consistent with predictions of dust evolution models of discs with and without strong pressure bumps \citep{Pinilla:2020,Pinilla:2025}, as well as predictions of dust evolution in photoevaporating discs \citep{Garate:2023}. However, it is unclear how this dust evolution process could be linked to the gas disc evolution.
Moreover, \citet{Marel:2021} found that structured discs are more commonly found around higher mass stars than lower mass stars, so if substructures result in longer disc lifetimes, longer disc lifetimes would be expected to be seen in higher mass stars rather than lower mass stars, which is the opposite of our findings.

On the other hand, enhanced radial drift could possibly result in a more rapid decrease in apparent disc lifetime. One scenario would be that, in the case of radial drift, all small grains are efficiently packed into large grains, which rapidly drift inwards and become lost by either sublimation by the star or very efficient planet formation through pebble accretion in the inner region of the disc \citep{Sanchez:2024}. If fragmentation is sufficiently limited so that small dust grains are not replenished \citep{Birnstiel:2010}, this might lead to more efficient photoevaporative clearing of the gas disc and thus a shorter gas disc lifetime. Investigating the infrared excess as a function of time using a dust evolution model that includes radial drift would allow one to determine whether a connection exists between dust and gas mass. If confirmed, the width of the disc lifetime distribution could be due to the degree of clustering in the disc. However, dust evolution models predict more efficient radial drift around lower mass stars \citep{Pinilla:2022}, so if this scenario holds, one would expect shorter disc lifetimes for lower mass stars, again the opposite of our findings. In summary, dust evolution processes are unlikely to explain the differences in disc lifetime as a function of stellar mass.

\subsection{Disc composition}
\label{sec:composition}

Variations in chemical composition may potentially lead to differences in the lifetimes of individual protoplanetary discs. Although understanding of the chemical composition of discs is still in an early stage, preliminary findings indicate a possible correlation between chemical composition and the mass of the host star. Discs around lower-mass stars ($T_{eff} <$~3000~K) are potentially (i) more carbon-rich and (ii) more molecule-rich than their higher-mass counterparts ($T_{eff} >$~4000~K) \citep{Walsh:2015}. The differences are especially pronounced in the inner disc regions \citep{Sellek:2025}. Observations of very low-mass stars ($<$~0.3~\MSun) show hydrocarbon-rich MIR spectra indicative of C/O~$>$~1 in their inner discs. 
In contrast, discs around higher-mass hosts are typically richer in O-bearing species. There, the molecules most frequently detected are H$_2$O, CO$_2$, C$_2$H$_2$, and HCN \citep[see,][and references therein]{Sellek:2025} with emissions mainly from within 1~au. However, there are exceptions, for example, DoAr~33 \citep{Colmenares:2024}, a 1.1~\MSun\ star with a hydrocarbon-rich chemistry consistent with C/O=2–4, and the water-rich inner disc of the very low-mass star Sz~114 \citep{Xie:2023}. However, recent modelling work points to dust evolution as an explanation for the diversity in composition rather than stellar mass \citep[e.g.][]{Krijt:2025}.

\section{Wide disc lifetime distributions}
\label{sec:width}

The derived disc lifetime distributions for all masses are relatively wide. Age spreads in clusters and associations contribute to the width of the disc lifetime distribution. However, it is likely not the dominant reason. As the work by \citet{Polnitzky:2026} shows, taking into account the different ages of the subclusters has only a very limited influence on the overall disc lifetime. Especially, age spreads can not account for the disc around 20~--~40~Myr old stars.

Next, we discuss possible reasons for the considerable variation in individual disc lifetimes of stars of similar masses.

\subsection{Variety in disc size, mass, and accretion rate}
\label{sec:width_mass}

The individual disc's lifetime could depend on its specific properties, such as size and mass. Larger discs generally have longer lifetimes because their expansive surface area can slow material dispersal. Similarly, a disc with substantial mass exerts a greater gravitational force, helping retain material for a longer period and delaying depletion. These properties, when considered together, illustrate how the size and mass of a disc can significantly influence its longevity. Observations have shown that the dust disc mass increases with the stellar mass approximately as $M_{dust} \approx (\Mstar)^{1.3-1.9}$ \citep{Pascucci:2016}. Nevertheless, individual disc masses and sizes vary by about three orders of magnitude for stars of similar mass \citep{Andrews:2020}. For example, gas disc masses can range from as low as 10~\MEarth to more than 1000~\MEarth, while gas disc sizes can vary from 50~au to 500~au in diameter. 
Thus, the broad distributions found for all mass bins may contribute to the variety in disc lifetimes.

Averaged over large numbers of discs, there exists a linear relation between the mass accretion rate onto the central star and the total mass of the disc \citep{Manara:2018}. However, again, the spread in the accretion rates is two to three orders of magnitude. Thus, in principle, high-mass discs with exceptionally low accretion rates and low-mass discs with untypical high-accretion rates could contribute to the spread in disc lifetimes. This could be tested by investigating distributions' outliers in detail.

However, care must be taken not to overinterpret these results. Most of these studies do not measure the gas mass but rather the mm-dust mass and use an ISM gas-to-dust ratio of $\sim$~100 to constrain the gas disc mass. mm-dust grains are known to be affected by transport (i.e. drift or trapping). However, if the mm-dust mass is controlled by dust evolution \citep{Marel:2021}, its value may not be representative of the gas disc mass and may not be used to constrain gas disc evolution.

\subsection{Environment}
\label{sec:width_env}

As most stars form in groups \citep{Lada:2003}, premature disc destruction can occur due to external photoevaporation \citep[e.g.,][]{Ercolano:2017} or stellar flybys \citep[e.g.,][]{Vincke:2016}. The relevance of these processes depends on the stellar density and the presence or absence of massive stars. The densities in young clusters vary by about seven orders of magnitude \citep{Pfalzner:2009}. Thus, the environment type could influence the disc fraction in individual clusters and therefore introduce additional width to the disc lifetime distribution.

Although environmental effects unquestionably can shorten the lifetimes of individual discs, they are unlikely to be the reason for the spread in disc lifetimes observed here. The data used in our analysis predominantly include environments with relatively low stellar density. Very high-density clusters where the environment may substantially influence the disc lifetime, such as the Arches or Quintuplet clusters, have been excluded to avoid contamination. The older clusters in our sample, especially Upp Sco and UCL/LCC, may have been denser in the past. However, they also contain some of the oldest disc-bearing stars. So, rather than spreading the disc lifetime to shorter time scales, they are hosting relatively long-lived discs. Thus, the environment may contribute but is a minor cause of the large spread in disc lifetimes.

\subsection{Two underlying distributions}
\label{sec:width_two}

\begin{figure}[ht]
\centering\includegraphics[width=0.49\textwidth]{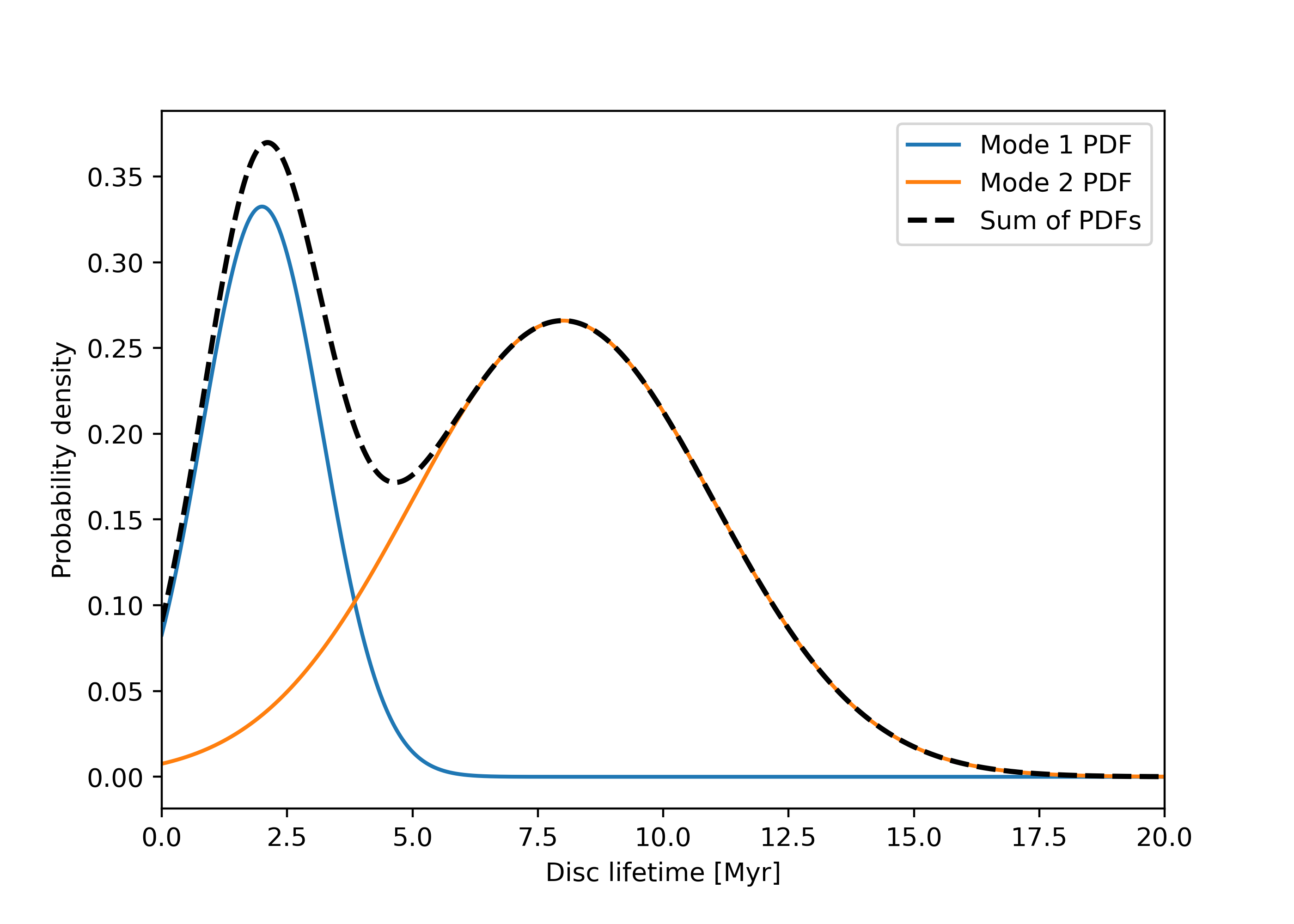}
\centering\includegraphics[width=0.44\textwidth]{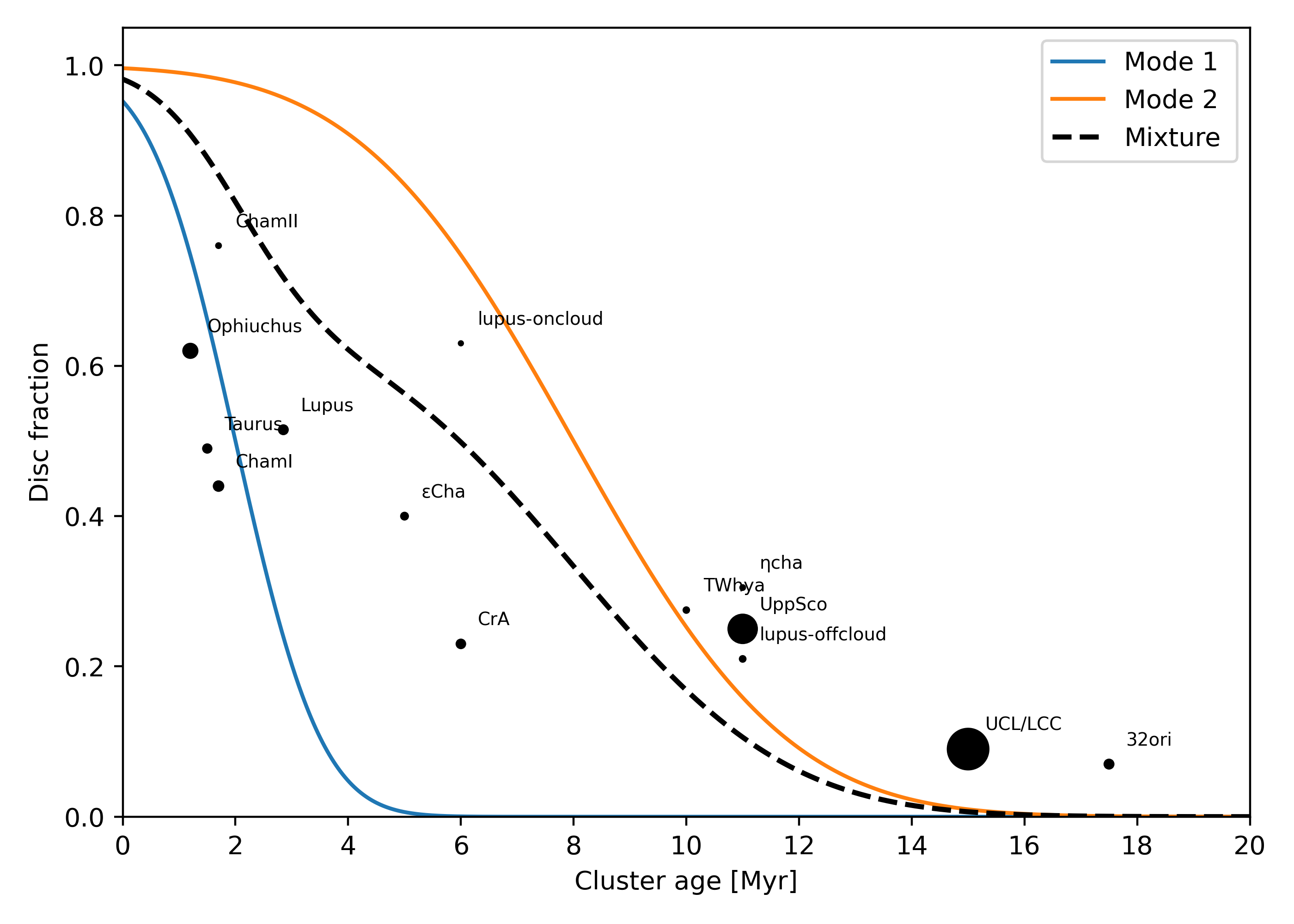}
\caption{Two types of discs with different disc lifetimes. Here, we assume a population of discs with a short (2.0~Myr; mode 1, blue line) and a long (8.0~Myr; mode 2, orange line) median disc lifetime. For simplicity, we assume that the distributions themselves are Gaussian (dashed black line). The top panel shows the distributions of these two populations and the resulting combined appearance of two such distributions. The bottom panel shows the corresponding decline in disc fraction as a function of cluster age.}
\label{fig:two_modes-1}
\end{figure}

Another reason for the large width of the disc lifetime distribution could be two underlying disc dispersal mechanisms operating on different timescales (see Fig.~\ref{fig:two_modes-1}). The wide distribution would result from the overlap of the two distinct distributions. The bimodal processes could be directly linked to the stellar mass, so that the widths and the mass-dependence are intrinsically linked. The left-weighted mass of the distribution results from combining two symmetric distributions. Several mechanisms could be responsible for this bimodality:

%Structured discs retained high dust masses, whereas the dust mass of the non-structured discs decreased over time (see Sect.~\ref{sec:structure}).
If discs with dust substructures indeed follow a separate dust evolutionary path than non-structured discs (see Sect.~\ref{sec:structure}) and the dust dynamics is connected in any way to the gas dynamics, the disc dispersal in structured and unstructured discs might be represented by separate disc lifetime distributions.

The dispersal of the bimodal discs could also be related to the composition of the disc. Discs seem to be correlated to the mass of the host star, with low-mass stars often being more carbon- and molecule-rich \citep{Arabhavi:2024}, while discs around higher mass stars show more O-bearing species \citep{Kaeufer:2026}. A bimodal distribution caused by the differences in disc chemistry would simultaneously explain the mass-dependence and the width of the disc-mass distribution. In such a scenario, carbon-rich and molecule-rich discs could have distinctly different disc dispersal timescales, with molecule-rich discs dispersed much more quickly than carbon-rich discs.

The mentioned exceptional cases would also allow for a prediction. Namely, molecule-rich discs around low-mass stars should be short-lived, and carbon-rich discs around high-mass stars should be long-lived. Thus, high-mass stars older than 6~Myr should preferentially be surrounded by carbon-rich discs. This prediction for a composition-based bimodal disc dispersal mechanism would be testable.

\section{Summary and Conclusion}
\label{sec:conclusion}

We determined the disc lifetime distribution of low- to higher-mass stars, with an emphasis on the high-mass end of the spectrum. We found that right-skewed distributions fit the observational data better than symmetric ones, not only for low-mass stars \citep{Pfalzner:2024}, but also for higher-mass stars. Assuming that initially 80~\% of stars were surrounded by a disc, a Weibull distribution of the form
\be
 T(x, k, \lambda) & = &  \frac{k}{6.1} \left(\frac{x}{6.1} \right)^{0.73} exp \left(- \left(\frac{x}{6.1} \right)^k \right)
\ee
well describes the distribution based solely on Upp Sco and UCL/LCC data. The median disc lifetime for high-mass stars (4.93~Myr) is significantly shorter than that for low-mass stars (8.52~Myr) assuming the same initial disc fraction of 80~\%. Enhanced radiation and stellar winds associated with high-mass stars are likely responsible for this reduced disc lifetime. The extent to which the composition and structure of discs contribute to accelerated disc dispersal remains unresolved.

We confirm previous findings that the initial disc fraction for star clusters is likely less than 100~\%, possibly $<$~80~\%. We find strong evidence that the initial disc fraction may also depend on stellar mass. For high-mass stars, it may be as low as $\sim$~40~\%. Possible reasons are the higher binarity of high-mass stars or stronger dynamical interactions in the cluster environment.

As for low-mass stars, the disc lifetime distribution of high-mass stars is fairly wide, with $\sigma =$~3.24~Myr for the above mass bins. Thus, individual disc lifetimes can differ significantly from the median (typical) lifetime. Differences in disc structure or composition could be responsible for this width. An underlying bimodal disc dispersion would be an alternative explanation. In addition, potentially, age spread in the population can contribute to the distribution's width.

This study should be regarded as a first step towards more detailed investigations. The main challenge is the lack of mass-dependent disc fractions. Nevertheless, the distributions derived here already provide more stringent constraints on disc dispersal theories than just mean disc lifetimes. Thus, it could be used to test theoretical models.

\bibliography{references}{}
\bibliographystyle{aa}

\appendix
\setcounter{figure}{0}    
\renewcommand\thefigure{A.\arabic{figure}}    
\setcounter{table}{0}    
\renewcommand\thetable{A.\arabic{table}}    

\begin{table*}[ht]
\caption{Disc fractions of star clusters closer than 200~au.}
\centering \begin{tabular}{lrclllcclccccc}
\hline
Identification & $d$ & Age & $N_{\star}$  & $f_d$ & Limit  & Median mass &  log $\rho_c$ &Source  \\
& pc  & Myr &              &           &    & [\MSun]& [\MSun/pc$^3$] \\
\hline
Alessi 13       & 108      & 30      &  162 & 0.049\footnote{possibly debris disc fraction} & 0.04 \MSun & &        & a)\\
UCL/LCC         & 150      & 15--20  & 3665 & 0.09$\pm$0.01                &                & 0.15 & -0.85--(-1.05) & b), c)\\
32 Ori          &  95      & 15--20  &  160 & 0.07$^{+0.03}_{-0.02}$       &                & 0.15 &                & d)\\
Upp Sco         & 145      & 10--12  & 1688 & 0.25$^{+0.05}_{-0.04}$           & 0.01 \MSun     & 0.15 & -0.59          & e), c)\\
Lupus-off cloud & 160      & 10--12  &  60  & 0.21 $\pm$ 0.06              & 0.05 \MSun     &      &                & e)\\
$\eta$ Cha      &  94      &  8--14  &  40  & 0.28$\pm$ 0.14/0.33$\pm$0.16 &                &      &                & f)\\
TW Hya          &  56      & 7--13   &  56  & 0.25/0.30, 0.19$^{+0.08}_{0.06}$ &            &      &                & f), g)\\
Lupus-on cloud  & 160      & 6       &  30  & 0.63 $\pm$ 0.04                  & 0.05 \MSun &      &                & e)\\
CrA             & 152      & 5       & 146  & 0.23 $\pm$ 0.4                   & 0.04 \MSun &      &                & f)\\
$\epsilon$ Cha  & 101      & 5 (3--8)&  90  & 0.5/0.3\footnote{disc fraction is much higher in the centre than the outskirts of $\eta$ Cha} &         &                                                             &                & h), f)\\
Lupus           & 158      & 2.6--3.1& 158  & 0.50/0.53                        & 0.03 \MSun &      &                & a), f)\\
Cham I          & 188      & 1.7     & 183  & 0.44                         & 6~$<$~G~$<$~20 &      &                & i)\\
Cham II         & 197      & 1.7     &  41  & 0.76                         & G12--G18       &      &                & i)\\
Taurus          & 128--196 & 1--2    & 137  & 0.49, 0.637                  & 0.05 \MSun     &      &                & j), k)\\
Ophiuchus       & 139      & 1--2    &  420 & 0.62                         &                &      &                & j)\\    
\hline  
\end{tabular}
\tablerefs{
a) \citet{Galli:2021_1},
b) \citet{Pecaut:2016},
c) \citet{Luhman:2021},
d) \citet{Luhman:2022},
e) \citet{Luhman:2020},
f) \citet{Michel:2021},
g) \citet{Luhman:2023},
h) \citet{Dickson:2021},
i) \citet{Galli:2021_2},
j) \citet{Manzo:2020},
k) \citet{Luhman:2023b}.}
\label{tab:cluster_properties}
\end{table*}  

\clearpage

\begin{figure}[ht]
 \centering
  \begin{minipage}[b]{0.48\textwidth}
    \centering
    \includegraphics[width=\textwidth]{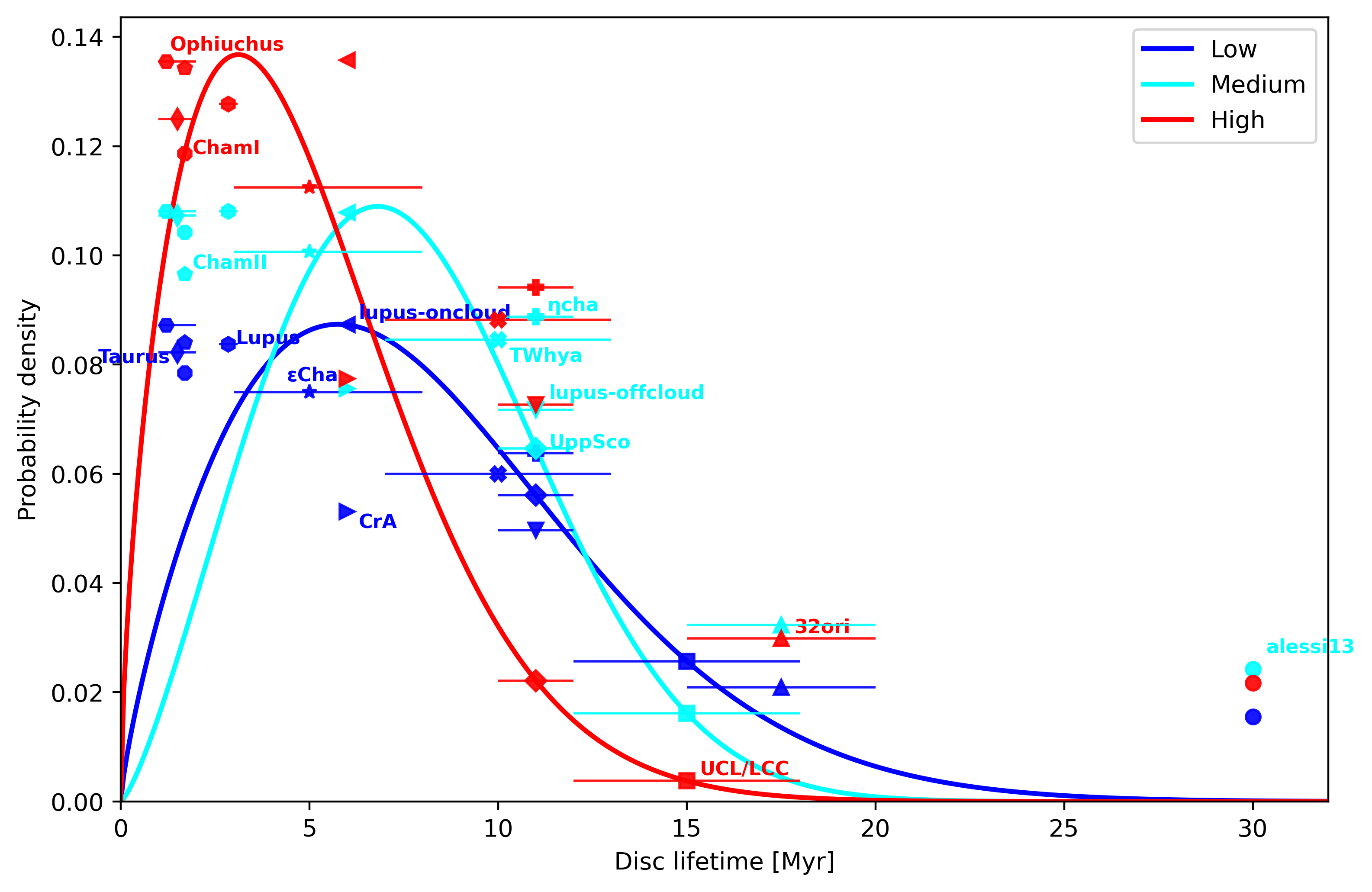}
    \textbf{(a)}
  \end{minipage}
  \begin{minipage}[b]{0.47\textwidth}
    \centering
    \includegraphics[width=\textwidth]{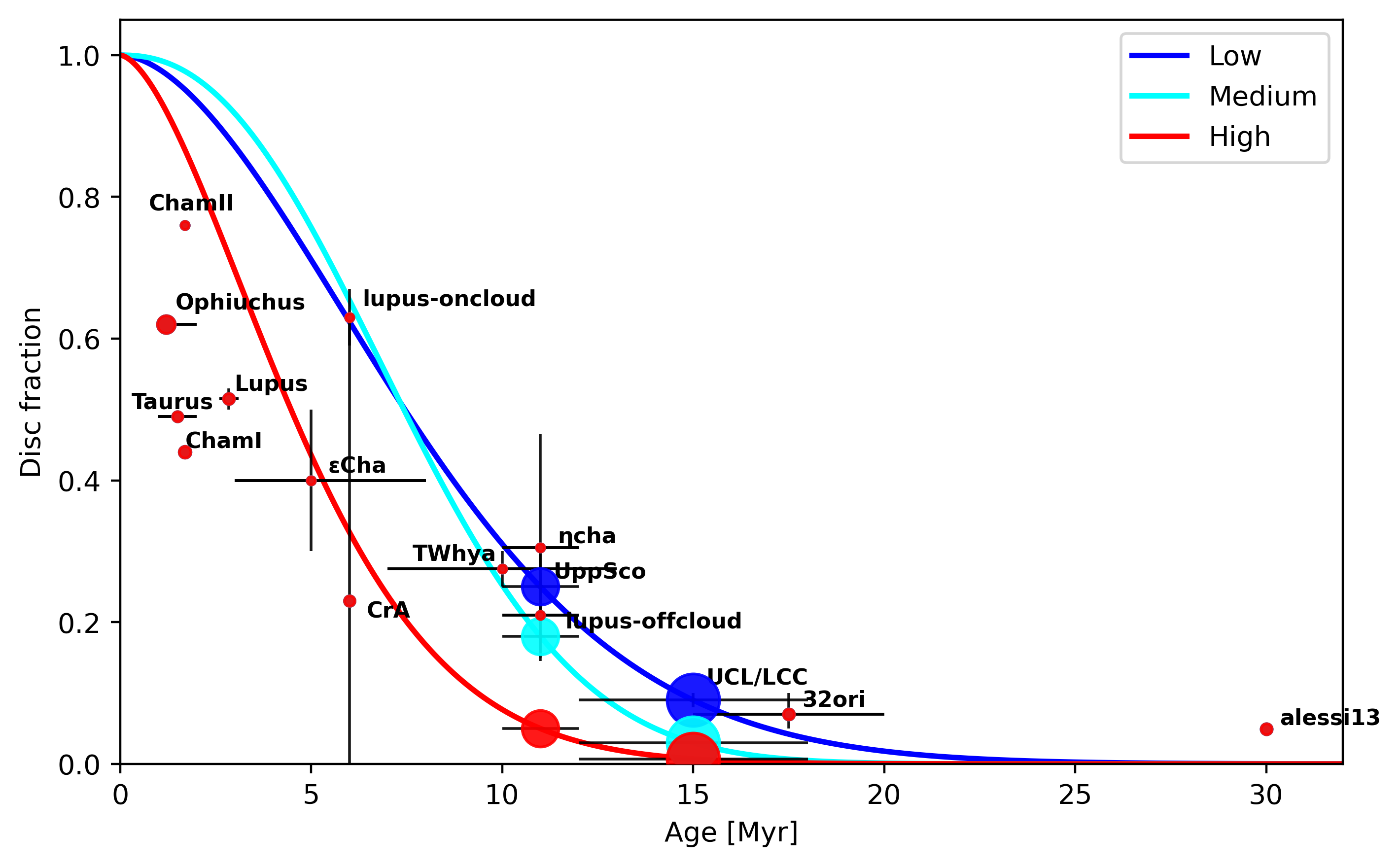}
    \textbf{(b)}
  \end{minipage}
\caption{Same as Fig.~\ref{fig:distribution_mass_100}, including the error bars for the star clusters.}
\label{fig:distribution_mass_error}
\end{figure}

\end{document}